\documentstyle[11pt]{article}
\def\bea{\begin{eqnarray}}
\def\eea{\end{eqnarray}}

\def\beq{\begin{equation}}
\def\eeq{\end{equation}}
\def\ba{\beq\new\begin{array}{c}}
\def\ea{\end{array}\eeq}
\def\be{\ba}
\def\ee{\ea}

\def\newpic#1{}

\makeatletter
\newdimen\normalarrayskip 
\newdimen\minarrayskip 
\normalarrayskip\baselineskip
\minarrayskip\jot
\newif\ifold \oldtrue \def\new{\oldfalse}
\def\arraymode{\ifold\relax\else\displaystyle\fi} 
\def\eqnumphantom{\phantom{(\theequation)}} 
\def\@arrayskip{\ifold\baselineskip\z@\lineskip\z@
\else
\baselineskip\minarrayskip\lineskip2\minarrayskip\fi}
\def\@arrayclassz{\ifcase \@lastchclass \@acolampacol \or
\@ampacol \or \or \or \@addamp \or
\@acolampacol \or \@firstampfalse \@acol \fi
\edef\@preamble{\@preamble
\ifcase \@chnum
\hfil$\relax\arraymode\@sharp$\hfil
\or $\relax\arraymode\@sharp$\hfil
\or \hfil$\relax\arraymode\@sharp$\fi}}
\def\@array[#1]#2{\setbox\@arstrutbox=\hbox{\vrule
height\arraystretch \ht\strutbox
depth\arraystretch \dp\strutbox
width\z@}\@mkpream{#2}\edef\@preamble{\halign
\noexpand\@halignto
\bgroup \tabskip\z@ \@arstrut \@preamble \tabskip\z@ \cr}%
\let\@startpbox\@@startpbox \let\@endpbox\@@endpbox
\if #1t\vtop \else \if#1b\vbox \else \vcenter \fi\fi
\bgroup \let\par\relax
\let\@sharp##\let\protect\relax
\@arrayskip\@preamble}
\def\eqnarray{\stepcounter{equation}%
\let\@currentlabel=\theequation
\global\@eqnswtrue
\global\@eqcnt\z@
\tabskip\@centering
\let\\=\@eqncr
$$%
\halign to \displaywidth\bgroup
\eqnumphantom\@eqnsel\hskip\@centering
$\displaystyle \tabskip\z@ {##}$%
\global\@eqcnt\@ne \hskip 2\arraycolsep
$\displaystyle\arraymode{##}$\hfil
\global\@eqcnt\tw@ \hskip 2\arraycolsep
$\displaystyle\tabskip\z@{##}$\hfil
\tabskip\@centering
&{##}\tabskip\z@\cr}
\begingroup\ifx\undefined\newsymbol \else\def\input#1 {\endgroup}\fi

\begin{document}

\setcounter{footnote}{1}
\def\thefootnote{\fnsymbol{footnote}}
\begin{center}
\hfill ITEP/TH-03/01\\
\hfill hep-th/0102069\\
\vspace{0.3in}
{\Large\bf Integrable Many-Body Systems via Inosemtsev Limit
}
\end{center}
\centerline{{\large Yu.Chernyakov}\footnote{
ITEP, Moscow, Russia; e-mail: chernyakov@gate.itep.ru}}
\centerline{{\large A.Zotov}\footnote{
ITEP, Moscow, Russia and MIPT, Dolgoprudny,
 Russia; e-mail: zotov@gate.itep.ru}}

\bigskip
\abstract{\footnotesize
The Inozemtsev limit (IL) or the scaling limit is known to be a
procedure applied to the elliptic Calogero Model.
It is a combination of the trigonometric limit, infinite shifts
of particles coordinates and rescalings of
the coupling constants.
As a result, one obtains an exponential type of interaction.
In the recent paper it is shown that the IL applied to
the $sl(N,\bf C)$ elliptic Euler-Calogero Model
and the elliptic Gaudin Model
produces new Toda-like systems of $N$ interacting particles
endowed with additional degrees of freedom corresponding to a
coadjoint orbit in $sl(n,\bf C)$. The limits corresponding
to the complete degeneration of the orbital degrees provide only
ordinary periodic and non periodic Toda systems.
We introduce a classification of the systems appearing in
the $sl(3,\bf C)$ case via IL.
The classification is represented on two-dimensional
space of parameters describing the infinite shifts of the coordinates.
This space is subdivided into symmetric domains.
The mixture of the Toda and the trigonometric Calogero-Sutherland potentials
emerges on the low dimensional domain walls of this picture.
Due to obvious symmetries this classification can be generalized to
the arbitrary number of particles.
We also apply IL to
$sl(2,\bf C)$ elliptic Gaudin Model with two marked points on the elliptic
curve and discuss main features of its possible limits.
The limits of Lax matrices are also considered.
}

\begin{center}
\rule{5cm}{1pt}
\end{center}

\bigskip
\setcounter{footnote}{0}
\section{Introduction.}

\paragraph{1. Inozemtsev limit}
Let us consider a system of two interacting particles with the
hamiltonian:
\be
H(v,u)=v^2+m^2E_2(u)
\ee
where $v$ and $-v$ their momenta in the
center of mass frame, $u$ is the difference between their
coordinates, $m$ is the coupling constant and $E_2(u,\tau)$ is the
Eisenstein function defined on the complex torus $T^2$
with moduli $\tau$. It is related to the Weierstrass functions
as follows:
$E_2(z,\tau)=\wp(z,\tau)+2\zeta(\frac{1}{2})$.

Inozemtsev limit (see [3],[4],[8])
is a combination of the trigonometric limit
$Im(\tau)\rightarrow\infty$ with the infinite shifts of coordinates
and a scaling of the coupling constant. This procedure can transform
the above hamiltonian into the following one written in the shifted
coordinates $U=u-\frac{1}{2}Im(\tau)$ and with the rescaled coupling
constant $\bar{m}$:
\be
\bar{H}=v^2+\bar{m}^2(e^U+e^{-U})
\ee
In a similar manner the IL may be applied to the system of $N$
interacting particles:
\be
H(v_i,u_i)=\sum\limits_{i=1}^{N}v_i^2+
m^2\sum\limits_{i\neq j}^{N}E_2(u_i-u_j)
\ee
The limit leads to the periodic Toda system
(non periodic case is also possible):
\be
H(v_i,U_i)=\sum\limits_{i=1}^{N}v_i^2+2
\bar{m}^2\sum\limits_{i,j=i+1;j=1,i=N}^{N}e^{U_j-U_i}
\ee
Thus IL describes the connection between
the Elliptic Calogero-Moser and the Toda models.

\paragraph{2. Purpose of the paper}
The purpose of the paper is to apply the IL to
different kinds of generalizations of the elliptic
Calogero model and to find out what sort of systems appear in
the limit depending on its parameters.

We are going to apply the IL to the following models:

$1.$ The $sl(N,\bf C)$ elliptic Euler-Calogero Moser model (ECM):
\be
L_{kl}(z,\bar{z})=v_k\delta_{kl}-p_{kk}\delta_{kl}E_1(z)
-p_{kl}(1-\delta_{kl})
e^{2\pi i\frac{z-\bar{z}}{\tau-\bar{\tau}}u_{kl}}\Phi(u_{kl},z)
\ee
where $p_{ij}\in sl(N,\bf C)$ and Poisson brackets are:
\be
\{v_i,u_j\}=\delta_{ij},
\ \
\{p_{ij},p_{kl}\}=p_{kj}\delta_{il}-p_{il}\delta_{kj},
\ \
\{p_{ij},v_k\}=0,
\ \
\{p_{ij},u_k\}=0
\ee
\be
H(u_i,v_i)=\sum\limits_{i=1}^{N}v_i^2+
\sum\limits_{i\neq j}p_{ij}p_{ji}E_2(u_{ij})
\ee
The integrable system is defined on the symplectic factor $sl(N,{\bf C})//
\{diag\ SL(N,\bf C)\}$.
In this case it is given by $p_{kk}=0$ and some gauge fixing condition
(see section $2$).
In the following the degrees of freedom related to $p_{ij}$
on the reduced space are called the orbital.

The hamiltonians $H_{k,l}$
are defined by the Lax matrix $L_{ij}(z,\bar{z})\in\ sl(N,\bf{C})$,
$\ z\in T^2$:
$$
Tr(L^k(z,\bar{z}))=\sum\limits_{l=0}^{N}E_{l}(z)H_{k,l}
$$
where $E_l(z)\sim\frac{1}{z^{l}}$ are the elliptic functions
(see Appendix A).

$2.$ The elliptic Gaudin model (EG) (see [6]):
\be
(L_{G})_{kl}=v_k\delta_{kl}-\sum\limits_{a=1}^{N_0}(p_a)_{kk}\delta_{kl}
E_1(z-x_a)-\\-
(1-\delta_{kl})\sum\limits_{a;k\neq l}e^{2\pi i
\frac{(z-x_a)-(\bar{z}-\bar{x}_a)}{\tau-\bar{\tau}}u_{kl}}(p_a)_{kl}
\Phi(u_{kl},z-x_a)
\ee
where $(p_a)_{kl}\in sl(N,\bf C)$. The Poisson brackets are:
\be
\{v_i,u_j\}=\delta_{ij},\ \ \
\{(p_a)_{ij},(p_b)_{kl}\}=\delta_{ab}((p_a)_{kj}\delta_{il}-
(p_a)_{il}\delta_{kj}),\\
\{(p_a)_{ij},u_k\}=0,\ \ \{(p_a)_{ij},v_k\}=0,\ \ \ a=1...N_0
\ee

\paragraph{3. Notations}
For  making  the limits  it is convenient to consider the torus
$T^2:
\bf C/(2\omega_1\bf{Z}+2\omega_2\bf Z)$,
$\ \omega_1=-i\pi,\ \ Im(\omega_2)=0,\  \tau=\frac{\omega_2}
{\omega_1}$.

The trigonometric limit is produced by $\omega_2\rightarrow \infty$.
Bar is used for the rescaled variables and function $\chi$ for
the powers of the exponents:
$m=\bar{m}e^{\chi_m\omega_2},\ \chi_m\equiv\chi(m)$.

The shifts of coordinates are introduced as follows:
\be
u_{ij}=U_{ij}+f_{ij}\omega_2
\ee

\paragraph{4. Results}
In the $sl(3,\bf C)$ ECM model the quadratic
hamiltonian follows from $(6)$:
$$
h_{2,0}=\sum\limits_{i=1}^{3}v_i^2-2aE_2(u_{12})-2bE_2(u_{23})-2cE_2(u_{31})
$$
where $a=p_{12}p_{21},\ b=p_{23}p_{32},\ c=p_{31}p_{13}$ are
the orbital variables
on the reduced phase space (the last one is $d=p_{12}p_{23}p_{31}$).

Let us write down the possible quadratic hamiltonians, obtained from the
above one via the IL, modulo permutations of the particles
and degenerations (some terms can vanish in the limit):
\be
1.\
\bar{h}_{2,0}=\sum\limits_{i=1}^{3}v_i^2-2\bar{a}e^{-U_{12}}-
2\bar{b}e^{-U_{23}}-2\bar{c}e^{-U_{31}}
\ee
\be
2.\
\bar{h}_{2,0}=\sum\limits_{i=1}^{3}v_i^2-2\bar{a}e^{U_{12}}-
2\bar{b}e^{-U_{23}}-2\bar{c}e^{-U_{31}}
\ee
\be
3.\
\bar{h}_{2,0}=\sum\limits_{i=1}^{3}v_i^2-
2\bar{a}(e^{-U_{12}}+e^{U_{12}})-
2\bar{b}e^{-U_{23}}-2\bar{c}e^{-U_{31}}
\ee
\be
4.\
\bar{h}_{2,0}=\sum\limits_{i=1}^{3}v_i^2-
\frac{1}{2}\bar{a}\frac{1}{\sinh^2{\frac{U_{12}}{2}}}-
2\bar{b}e^{-U_{23}}-2\bar{c}e^{U_{31}}
\ee
\be
5.\
\bar{h}_{2,0}=\sum\limits_{i=1}^{3}v_i^2-
\frac{1}{2}\bar{a}\frac{1}{\sinh^2{\frac{U_{12}}{2}}}-
2\bar{b}(e^{-U_{23}}+e^{U_{23}})-2\bar{c}(e^{U_{31}}+e^{-U_{31}})
\ee
\be
6.\
\bar{h}_{2,0}=\sum\limits_{i=1}^{3}v_i^2-
\frac{1}{2}\bar{a}\frac{1}{\sinh^2{\frac{U_{12}}{2}}}-
\frac{1}{2}\bar{b}\frac{1}{\sinh^2{\frac{U_{23}}{2}}}-
\frac{1}{2}\bar{c}\frac{1}{\sinh^2{\frac{U_{31}}{2}}}
\ee
The first and the last one are a direct generalizations of the periodic
Toda and the Sutherland-Calogero systems correspondingly.
All others content an unusual dependence on the coordinates.

The classification can be given by the following figure
($f_{ij}$ are parameters of the shifts of coordinates: $u_{ij}=
U_{ij}+\omega_2 f_{ij}$ and thus $f_{ij}+f_{jk}+f_{ki}=0$):
\vskip 10mm
\begin{picture}(350,200)(-20,-10)
\put(310,0){$\hbox{fig.1}$}
\bezier{500}(10,3)(70,105)(130,207)
\bezier{500}(110,3)(170,105)(180,122)
\bezier{500}(15,20)(100,20)(240,20)
\bezier{500}(110,190)(120,190)(140,190)
\bezier{500}(60,105)(100,105)(190,105)
\bezier{500}(210,3)(230,37)(230,37)
\bezier{500}(10,37)(20,20)(30,3)
\bezier{500}(130,3)(110,37)(60,122)
\bezier{500}(230,3)(220,20)(110,207)

\put(30,-5){$f_{31}=0$}
\put(130,-5){$f_{31}=-1$}
\put(230,-5){$f_{31}=-2$}
\put(-18,16){$f_{12}=0$}
\put(27,101){$f_{12}=1$}
\put(77,186){$f_{12}=2$}
\put(133,205){$f_{23}=0$}
\put(183,120){$f_{23}=1$}
\put(233,40){$f_{23}=2$}

\put(70,48){$II$}
\put(170,48){$IV$}
\put(120,133){$III$}
\put(120,77){$I$}
\end{picture}

On fig .1 different kinds of the systems correspond to symplexes of
different dimensions and situations (modulo permutations of $a$,$b$,$c$):
$\{I\}$ and $\{III\}$ - two types of two dimensional symplexes,
which content the limits to hamiltonians of the first
$(11)$ and the second $(12)$ types (domains $\{II\}$ and $\{IV\}$
are equivalent to $\{III\}$ in sense of the permutations);
lines $\{f_{ij}=2\bf Z\}$ and midlines $\{f_{ij}=1+2\bf Z\}$ content
intervals which correspond to the mixture of the
Sutherland and the Toda $(14)$ and
$\cosh(U_{ij})$ terms $(13)$;
all vertices - zero dimensional symplexes correspond to fifth $(15)$ and
sixth $(16)$ types of the hamiltonians.

The considering of the cubic hamiltonians provides more complicated
structure on the fig.$1$. by adding a dual lattice
(see fig.$2$ in Section $4$).

The resultant picture (unification of fig.$1$ and fig.$2$)
appear to describe the classification of
possible limits of brackets between $\bar{a}$, $\bar{b}$, $\bar{c}$
and $\bar{d}$.

Fig.$1$,$2$ have obvious symmetries and thus the classification can be
generalized to an arbitrary number of particles. In the case of $N$
interacting particles all possible limits of hamiltonians and
Poisson brackets correspond to different types of domains in the
$N-1$ dimensional hyperplane of variables $f_{ij}$.

There are more possible limits in the EG because the diagonal part of
the Lax matrix $(9)$
content dependence on $(p_a)_{ii}$ even on the reduced phase space
and there are additional parameters $(x_a-x_b)$
which can be also shifted in the limit. In this case
resultant hamiltonians
content a direct dependence on the parameters of the limits
(see Section $7$).

\section{EG and $sl(N,\bf C)$ ECM.}
The Elliptic Gaudin Model is an example of the Hitchin system [2]
(see also [5] and [7]) and was introduced in [6].
The Lax matrix is given by $(8)$.

We can restrict algebra $(9)$ on the submanifold:
\be
\sum\limits_{a}(p_a)_{kk}=0,\ \ k=1...N
\ee
since it corresponds to the fixing of the
moment map $\mu_p$ by the adjoint action of the
diagonal $SL(N,\bf{C})$ subgroup on $p_{kl}$ and the
fact $p_{kk}$ are Poisson commute with the $Tr(L(z)^n)$.
By fixing the adjoint action one reduces the number of independent
variables in $(p_a)_{kl}$ by $N-1$.
Note that the restriction $(17)$ is in agreement with
the simple requirement on holomorphic function:
\be
\sum\limits_{simple\ poles} res\{(L_{G}(z))_{kk}\}=0
\ee
The hamiltonians appear as the coefficients in the
equation describing the spectral curve:
\be
det(L(z)+\mu)=0
\ee

Another way is to decompose  $Tr(L^m)$  on the basis of functions
$\{E_1(z-x_a),\ E_2(z-x_a),\ E_2'(z-x_a),...\}$.
Then the coefficients in the decomposition are again hamiltonians
(see [6],[5]).

The $sl(N,\bf C)$ ECM corresponds to $N_0=1$. In this case
$(17)$ is written as follows:
\be
(\mu_p)_i=p_{ii}=0\ \ i=1...N
\ee
and thus the Lax matrix is:
\be
(L)_{kl}=v_k\delta_{kl}-\sum\limits_{k\neq l}e^{2\pi i
\frac{z-\bar{z}}{\tau-\bar{\tau}}u_{kl}}p_{kl}
\Phi(u_{kl},z)
\ee
Using appendix B we obtain to the following hamiltonians in this case:
\be
<L(z)^2>=H_{2,0}+H_{2,2}E_2(z),
\ee
\be
<L(z)^3>=H_{3,0}+H_{3,2}E_2(z)+H_{3,3}E_2'(z),
\ee
\be
H_{2,0}=\sum\limits_{i,j=1}^{N}({v_i}^2-p_{ij}p_{ji}E_2(u_{ij})),
\ee
\be
H_{2,2}=\sum\limits_{i,j=1}^{N}p_{ij}p_{ji},
\ee
\be
H_{3,0}=\sum\limits_{i,j,k=1}^{N}({v_i}^3-3v_ip_{ij}p_{ji}E_2(u_{ij})-
p_{ij}p_{jk}p_{ki}[E_2(u_{ij})E_1(u_{jk})+E_2(u_{jk})E_1(u_{ij})+\\
+\frac{1}{2}E_1(u_{ki})(E_2(u_{ij})+E_2(u_{jk}))+
\frac{1}{2}E_1(u_{ij}-u_{jk})
(E_2(u_{ij})-E_2(u_{jk}))]),
\ee
\be
H_{3,2}=\sum\limits_{i,j=1}^{N}(3v_ip_{ij}p_{ji}+p_{ij}p_{jk}p_{ki}
[E_1(u_{ij})+E_1(u_{jk})+E_1(u_{ki})]),
\ee
\be
H_{3,3}=-\frac{1}{2}\sum\limits_{i,j=1}^{N}p_{ij}p_{jk}p_{ki}.
\ee
Note that $H_{2,2}$ and $H_{3,3}$ are the Casimir functions of the
algebra $(6)$.

It should be also mentioned that the exponential factors in the Lax
matrices $(5),\ (8)$ are not necessary for computing the hamiltonians
since $u_{ij}+u_{ji}=0,\ u_{ij}+u_{jk}+u_{ki}=0$.
However these factors plays an important role in the limits of
the Lax matrix itself (see Section $8$).

\section{$sl(3,\bf C)$ ECM.}
Let us compute the dimension of the phase space of the $sl(N,\bf C)$ ECM.
In the $p_{ij}$ space a number of the independent variables equals
$(N-1)(N-2)$ since the integrable system is defined on the orbit
$\{p_{ij}//Diag\ SL(N,\bf C)\}$ with the values of the Casimir functions
fixed, where $Diag SL(N,\bf C)$ is the Cartan subgroup in the $SL(N,\bf C)$.
In the space $(u,v)\in T^*Bun_{G,\Sigma}$ there are $2(N-1)$
such variables. Thus the total dimension of the phase space equals
$N(N-1)$ and consequently $dim(sl(3,{\bf C})\ ECM)=6$ that is in the
agreement with a fact there are three integrals of motion in this case:
$H_{2,0}$, $H_{3,0}$, $H_{3,2}$.

The orbit $\{p_{ij}//Diag\ SL(N,\bf C)\}$ is given by conditions
$p_{ii}=0$ and some fixation of the coadjoint action by the Cartan subgroup.
However a certain fixation of the coadjoin action (either a fixation
of values of the Casimir functions) may be not useful. Thus we will
only imply
some choice of the gauge fixing condition without any exact one when
consider not invariant quantities.

Poisson algebra $(6)$ between nondiagonal variables in $N=3,N_0=1$
case looks like:
\be
\{p_{12},p_{23}\}=-p_{13},\ \ \{p_{32},p_{21}\}=-p_{31},
\\
\{p_{31},p_{12}\}=-p_{32},\ \ \{p_{21},p_{13}\}=-p_{23},
\\
\{p_{23},p_{31}\}=-p_{21},\ \ \{p_{13},p_{32}\}=-p_{12}.
\ee
We may do not care about fixing the action of $Diag\ SL(N,\bf C)$
if choose variables which are invariant with respect to it:
\be
a=p_{12}p_{21},\ b=p_{23}p_{32},\ c=p_{13}p_{31},\ d=p_{12}p_{23}p_{31}
\ee
Note that  $a$, $b$, $c$ and $d$ are the combinations of $p_{kl}$
sufficient for introducing the hamiltonians $(24-28)$.
From $(29)$ we obtain the brackets between the invariant variables:
\be
\{a,b\}=\{b,c\}=\{c,a\}=d-\frac{abc}{d},\\
\{a,d\}=ab-ac,\ \ \{b,d\}=bc-ab,\ \ \{c,d\}=ac-cb
\ee
Note also that there is no singularity in $\frac{abc}{d}$
since it equal to $p_{21}p_{32}p_{13}$.
Hamiltonians are given by $(24)-(28)$:
\be
h_{2,0}=v_1^2+v_2^2+v_3^2-2aE_2(u_{12})-2bE_2(u_{23})-2cE_2(u_{31})
\ee
\be
h_{2,2}=2(a+b+c)
\ee
\be
h_{3,0}=v_1^3+v_2^3+v_3^3
+3v_3aE_2(u_{12})+3v_1bE_2(u_{23})+3v_2cE_2(u_{31})+\\
+\frac{3}{2}(d-\frac{abc}{d})
\frac{E_2'(u_{12})E_2(u_{23})-E_2'(u_{23})E_2(u_{12})}
{E_2(u_{12})-E_2(u_{23})}
\ee
\be
h_{3,2}=3(v_1+v_2)a+3(v_1+v_3)c+
3(v_2+v_3)b+\\+3(d-\frac{abc}{d})
[E_1(u_{12})+E_1(u_{23})+E_1(u_{31})]
\ee
\be
h_{3,3}=-\frac{3}{2}(d+\frac{abc}{d})
\ee
Here $h_{2,2}$ and $h_{3,3}$ are the quadratic and the cubic Casimir
functions of the algebra $(31)$.

The spectral curve is written as follows:

\be
det(L(z)+\mu)=0\\
\mu^3-\frac{1}{2}\mu(E_2(z)h_{2,2}+h_{2,0})
+\frac{1}{3}(E_2'(z)h_{3,3}+E_2(z)h_{3,2}+h_{3,0})=0
\ee
\section{Degenerations of elliptic functions.}
\paragraph{1. Formulas used for computing IL in quadratic hamiltonians}

\be
\hbox{Let us put}
\ x=x_n+\sigma\omega_2,\ \hbox{where}\ x_n\ \hbox{is a new (shifted)
coordinate},\\
\ |\sigma|<2,\ \ \ 0<\beta<1,\ \ \
\omega_2\rightarrow\infty
\ee
Below we right down the main non vanishing order in the limit
$\omega_2\rightarrow\infty$.
\be
E_1(x)=\frac{1}{2}\sum\limits_{k=-\infty}^{\infty}
\coth(\frac{x}{2}-k\omega_2)\approx
\left\{
\begin{array}{l}
1.\ \sigma=0:\ \frac{1}{2}\coth(\frac{x_n}{2})\\
2.\ 0<\sigma<1:\ \frac{1}{2}+e^{-x_n-\sigma\omega_2}\\
3.\ \sigma=1:\ \frac{1}{2}-2e^{-\omega_2}\sinh(x_n)\\
4.\ \sigma>1:\ \frac{1}{2}-e^{x_n-(2-\sigma)\omega_2}\\
\end{array}
\right.
\ee

$$
e^{\beta\omega_2}E_2(x)=\sum\limits_{k=\infty}^{\infty}
\frac{e^{\beta\omega_2}}{e^{x+2k\omega_2}+e^{-x-2k\omega_2}-2}
\approx
$$
$$
\approx\frac{e^{\beta\omega_2}}{e^{x_n+\sigma\omega_2}+
e^{-x_n-\sigma\omega_2}-2}+\frac{e^{\beta\omega_2}}{e^{x_n+
\sigma\omega_2-2\omega_2}+e^{-x_n-\sigma\omega_2+2\omega_2}-2}\approx
$$
\be
\\
\approx
\left\{
\begin{array}{l}
\ \ 1.\ \ \beta=0,\ 0<\sigma<2
\\
1.1\ \ \sigma=0,\ \beta=0:\ \frac{1}{4}\frac{1}{\sinh^2{\frac{x_n}{2}}}
\\
1.2\ \ \beta=0,\ 0<\sigma<1:\ e^{-\sigma\omega_2-x_n}
\\
1.3\ \ \beta=0,\ \sigma=1:\ e^{-\omega_2}(e^{-x_n}+e^{x_n})
\\
1.4\ \ \beta=0,\ 1<\sigma<2:\ e^{-(2-\sigma)\omega_2+x_n}
\\
\ \ 2.\ \ \beta>0,\ 0<\sigma<2,\:
\\
2.1.\ \ \sigma=\beta< 2-\sigma:\ e^{-x_n}
\\
2.2.\ \ 2-\sigma=\beta<\sigma:\ e^{x_n}
\\
2.3\ \ \sigma=\beta=2-\sigma:\ e^{-x_n}+e^{x_n}
\end{array}
\right.
\ee
From $(40)$ one can simply obtain the quadratic hamiltonians
$(11)-(16)$ and $(4)$ given in the Introduction. For example
$(4)$ appears if one put
$
u_j=U_j+2j\omega_2\frac{1}{N}\ \
m=\bar{m}e^{\frac{1}{N}\omega_2}
$
(see [8]).
\paragraph{2. Formulas used for computing IL in cubic hamiltonians}
\be
E_2'(x)=-\frac{1}{4}\sum\limits_{k=-\infty}
\frac{\cosh(\frac{x}{2}-k\omega_2)}{\sinh^3(\frac{x}{2}-k\omega_2)}
\approx
\left\{
\begin{array}{c}
1.\ \ 0<\sigma<1:\ -e^{-\sigma\omega_2-x_n}\\
2.\ \ \sigma=1:\ e^{-\omega_2}(e^{x_n}-e^{-x_n})\\
3.\ \ 1<\sigma<2:\ e^{-(2-\sigma)\omega_2+x_n}\\
\end{array}
\right.
\ee
Let us consider the limits of following expression:
\be
\triangle=
-\frac{1}{2}\frac{E_2'(u_{12})E_2(u_{23})-E_2'(u_{23})E_2(u_{12})}
{E_2(u_{12})-E_2(u_{23})}
\ee
It is contented in $h_{3,0}$ given by $(34)$.
To analyze this expression one should
take into account terms of the second non vanishing
order in the approximate
formulas for $E_2(z)$ and $E_2'(z)$. The result is represented on fig.$2$:
\vskip 10mm
\begin{picture}(350,200)(-20,0)
\put(340,5){$\hbox{fig.2}$}
\bezier{500}(10,3)(70,105)(130,207)
\bezier{500}(110,3)(170,105)(230,207)
\bezier{500}(15,20)(100,20)(300,20)
\bezier{500}(110,190)(120,190)(240,190)
\bezier{500}(60,105)(100,105)(290,105)
\bezier{500}(210,3)(230,37)(280,122)
\bezier{500}(10,37)(20,20)(30,3)
\bezier{500}(130,3)(110,37)(60,122)
\bezier{500}(230,3)(220,20)(110,207)
\bezier{500}(210,207)(220,190)(280,88)
\bezier{500}(120,77)(120,100)(120,133)
\bezier{500}(220,77)(220,100)(220,133)
\bezier{500}(70,48)(70,20)(70,15)
\bezier{500}(170,48)(170,20)(170,15)
\bezier{500}(170,195)(170,190)(170,162)
\bezier{500}(120,133)(120,133)(170,162)
\bezier{500}(170,162)(170,162)(220,133)
\bezier{500}(120,77)(120,77)(170,48)
\bezier{500}(170,48)(170,48)(220,77)
\bezier{500}(120,133)(120,133)(87,152)
\bezier{500}(220,133)(220,133)(253,152)
\bezier{500}(220,77)(220,77)(262,58)
\bezier{500}(70,48)(70,48)(120,77)
\bezier{500}(70,48)(70,48)(30,67)
\put(30,5){$f_{31}=0$}
\put(130,5){$f_{31}=-2$}
\put(230,5){$f_{31}=-4$}
\put(-18,16){$f_{12}=0$}
\put(27,101){$f_{12}=2$}
\put(77,186){$f_{12}=4$}
\put(133,200){$f_{23}=0$}
\put(233,200){$f_{23}=2$}
\put(283,125){$f_{23}=4$}
\put(45,32){$C_{+}$}
\put(195,117){$C'_{+}$}
\put(135,82){$C_{-}$}
\put(95,82){$B_{-}$}
\put(195,82){$B'_{-}$}
\put(135,117){$B'_{+}$}
\put(85,32){$B_{+}$}
\put(65,70){$A_{+}$}
\put(165,70){$A'_{+}$}
\put(115,50){$A_{-}$}
\put(165,135){$A'_{-}$}
\end{picture}
\vskip 10mm
\be
\\
-\frac{1}{2}\frac{E_2'(u_{12})E_2(u_{23})-E_2'(u_{23})E_2(u_{12})}
{E_2(u_{12})-E_2(u_{23})}\approx
\left\{
\begin{array}{c}
A_{+}:e^{U_{12}}e^{-(2-f_{12})\omega_2}\\
A_{-}:-e^{-U_{12}}e^{-f_{12}\omega_2}\\
B_{+}:e^{U_{23}}e^{-(2-f_{23})\omega_2}\\
B_{-}:-e^{-U_{23}}e^{-f_{23}\omega_2}\\
C_{+}:e^{U_{31}}e^{f_{31}\omega_2}\\
C_{-}:-e^{-U_{31}}e^{-(2+f_{31})\omega_2}\\
\end{array}
\right.
\ee

Areas $A'_{-}$, $A'_{+}$, $B'_{-}$, $B'_{+}$ and $C'_{-}$
are shifted on the hole period with respect to
$A_{-}$, $A_{+}$, $B_{-}$, $B_{+}$ and $C_{-}$ correspondingly.

There are two types of boundaries on fig.$2$. The first corresponds to the
case when the boundary belongs to one of the axis ($f_{12}\in 2\bf Z$
or $f_{23}\in 2\bf Z$ or $f_{31}\in 2\bf Z$).
All other lines are contented in the second type. They belong to the dual
lattice.
In fig.$2$ one can see a hexagon $A'_{+}B'_{-}C'_{+}A'_{-}B'_{+}C_{-}$.
The sides of this hexagon content the boundaries of the first type and
the lines coming through the centers of the hexagons content the
boundaries of the second type.
Let us write down an example of limits of $\triangle$
corresponding to the first type
of the boundaries:
\be
\triangle\approx
\left\{
\begin{array}{l}
C_{+}\cap \{f_{23}=0\}:-e^{-U_{12}}e^{-f_{12}}
(1-\coth\frac{U_{23}}{2})\\
A_{+}\cap \{f_{23}=0\}:e^{U_{12}}e^{-(2-f_{12})}
(1+\coth\frac{U_{23}}{2})
\end{array}
\right.
\ee
The boundaries of the first type correspond to a combination
of the trigonometric limit and the IL and thus
content the mixture of the Toda and the Calogero-Sutherland potentials.

As for the boundaries of the second type the values of
$\triangle$ from the neighboring
domains are of the same order and thus the potential is equal to the sum
of those from the neighboring domains.
(All the same is valid for the zero dimensional boundary.)

\paragraph{3. $sl(N,\bf C)$ case}

In the $N$ particle case one should consider $tr(L^2)$,...,$tr(L^N)$.
The most complicated potential appears as a free term in the expression:
\be
\Phi(u_{12},z)\Phi(u_{23},z)...\Phi(u_{N1},z)
\ee

It is possible to generalize fig.$2$ on higher dimensions.
In $sl(4,\bf C)$ case the space of the parameters is three dimensional
($f_{12}+f_{23}+f_{34}+f_{41}=0$).
The whole space is subdivided into regular tetrahedrons with a
three dimensional structure inside that correspond to the
fig.$2$ on the two dimensional walls.
The case of the arbitrary $N$ can be obtained by analogy.

\section{Limits of brackets for $sl(3,\bf C)$ case.}
We are going to find non-singular limits of the brackets $(31)$ and
all possible limits of the hamiltonians $(32-36)$ for each of them.

The idea is to multiply if it is necessary a certain  hamiltonian
by the exponent with the power canceling the highest power in the
hamiltonian.
\be
h_{i,j}=\bar{h}_{i,j}e^{\chi(h_{i,j})}, \ \ \chi(h_{i,j})\geq 0
\ee

This procedure saves the involution property of the hamiltonians but
certainly does
not guaranty the independence of $\bar{h}_{i,j}$. The sufficient condition
for the independence of the hamiltonians is a saving of the kinetic terms
$\sum v_i^2$ and $\sum v_i^3$ in $h_{2,0}$ and $h_{3,0}$ correspondingly.

Another (equivalent to the previous) way of studding the limits
comes from considering the spectral curve $(37)$.
$$
\mu^3-\frac{1}{2}\mu(E_2(z)h_{2,2}+h_{2,0})
+\frac{1}{3}(E_2'(z)h_{3,3}+E_2(z)h_{3,2}+h_{3,0})=0
$$
One may require to have a non singular spectral curve in the limit.
We will see it is true for all limits if $\chi(h_{3,0})=\chi(h_{2,0})=0$.

One can also make the rescaling of the spectral parameter:
\be
\mu=\bar{\mu}e^{\chi_\mu}
\ee
However in this case the kinetic terms vanish in $h_{2,0}$ and $h_{3,0}$
and this may lead to the lost of independence of the hamiltonians.
To save the independence one may resale momenta
\be
v_i=\bar{v}_ie^{\chi_{\mu}}
\ee
But this can be done only if there is a factor $e^{-\chi_{\mu}}$ in
the canonical form:
\be
\omega=e^{-\omega_2\chi_{\mu}}\sum\limits_{i=1}^{3}dv_i\wedge du_i
\ee
In this sense the rescaling of the momenta does not provide interesting
consequences. We consider the case $\chi_{\mu}=0,\ \chi(h_{2,0})=0,\
\chi(h_{3,0})=0$.

The purpose of the section is to find possible (non-singular)
limits of the brackets $(31)$:
$$
\{a,b\}=\{b,c\}=\{c,a\}=d-\frac{abc}{d},\ \
\{a,d\}=ab-ac,\ \ \{b,d\}=bc-ab,\ \ \{c,d\}=ac-cb.
$$

and thus find out whether there exist limits with
non-trivial dynamics of rescaled $a,b,c,d$:
\be
\bar{a}=e^{\chi_a\omega_2}a,\ \ \ \ \bar{b}=e^{\chi_b\omega_2}b,\ \ \ \
\bar{c}=e^{\chi_c\omega_2}c,\ \ \ \ \bar{d}=e^{\chi_d\omega_2}d.
\ee

The non singularity conditions for  the rescaling of the
brackets $(31)$ are:

\be
\chi_a\leq\chi_d,\ \ \ \ \chi_b\leq\chi_d,\ \ \ \ \chi_c\leq\chi_d,\\
\chi_d\leq\chi_a+\chi_c,\ \ \ \chi_d\leq\chi_a+\chi_b,\ \ \
\chi_d\leq\chi_b+\chi_c.
\ee

It is important to notice that it is impossible to satisfy the
above non-equalities if some $\chi<0$.

Non triviality of the rescaled brackets means an exact equality
besides any of the inequalities $(51)$. Let us write down all
such possibilities modulo permutations of $a,\ b,\ c$:

\be
1:
\left\{
\begin{array}{l}
\chi_a<\chi_d,\ \chi_b<\chi_d,\ \chi_c<\chi_d,\ \chi_a+\chi_b=\chi_d,\
\chi_a+\chi_c>\chi_d,\ \chi_b+\chi_c>\chi_d\\
\{\bar{a},\bar{b}\}=\bar{d},\ \{\bar{b},\bar{c}\}=0,\ \{\bar{c},\bar{a}\}=0
,\ \{\bar{a},\bar{d}\}=0,\ \ \{\bar{b},\bar{d}\}=0,
\ \ \{\bar{c},\bar{d}\}=0
\end{array}
\right.
\ee
\be
2:
\left\{
\begin{array}{l}
\chi_a<\chi_d,\ \chi_b<\chi_d,\ \chi_c<\chi_d,\ \chi_a+\chi_b=\chi_d,\
\chi_a+\chi_c=\chi_d,\ \chi_b+\chi_c>\chi_d\\
\{\bar{a},\bar{b}\}=\bar{d},\ \{\bar{b},\bar{c}\}=\bar{d},
\ \{\bar{c},\bar{a}\}=0,\
\{\bar{a},\bar{d}\}=0,\ \ \{\bar{b},\bar{d}\}=0,
\ \ \{\bar{c},\bar{d}\}=0
\end{array}
\right.
\ee
\be
3:
\left\{
\begin{array}{l}
\chi_a<\chi_d,\ \chi_b<\chi_d,\ \chi_c<\chi_d,\ \chi_a+\chi_b=\chi_d,\
\chi_a+\chi_c=\chi_d,\ \chi_b+\chi_c=\chi_d\\
\{\bar{a},\bar{b}\}=\bar{d},\ \{\bar{b},\bar{c}\}=\bar{d},
\ \{\bar{c},\bar{a}\}=\bar{d},\
\{\bar{a},\bar{d}\}=0,\ \ \{\bar{b},\bar{d}\}=0,
\ \ \{\bar{c},\bar{d}\}=0
\end{array}
\right.
\ee
\be
4:
\left\{
\begin{array}{l}
\chi_a=\chi_d,\ \chi_b<\chi_d,\ \chi_c<\chi_d,\ \chi_a+\chi_b>\chi_d,\
\chi_a+\chi_c>\chi_d,\ \chi_b+\chi_c>\chi_d\\
\{\bar{a},\bar{b}\}=0,\ \{\bar{b},\bar{c}\}=
-\frac{\bar{a}\bar{b}\bar{c}}{\bar{d}},\ \{\bar{c},\bar{a}\}=
\{\bar{a},\bar{d}\}=0,\ \ \{\bar{b},\bar{d}\}=-\bar{a}\bar{b},
\ \ \{\bar{c},\bar{d}\}=\bar{a}\bar{c}
\end{array}
\right.
\ee
\be
5:
\left\{
\begin{array}{l}
\chi_a=\chi_d,\ \chi_b=\chi_d,\ \chi_c<\chi_d,\ \chi_a+\chi_b>\chi_d,\
\chi_a+\chi_c>\chi_d,\ \chi_b+\chi_c>\chi_d\\
\{\bar{a},\bar{b}\}=0,\ \{\bar{b},\bar{c}\}=
\{\bar{c},\bar{a}\}=-\frac{\bar{a}\bar{b}\bar{c}}{\bar{d}},\\
\{\bar{a},\bar{d}\}=\bar{a}\bar{b},
\ \ \{\bar{b},\bar{d}\}=-\bar{a}\bar{b},
\ \ \{\bar{c},\bar{d}\}=\bar{a}\bar{c}-\bar{c}\bar{b}
\end{array}
\right.
\ee
\be
6:
\left\{
\begin{array}{l}
\chi_a=\chi_d,\ \chi_d=\chi_d,\ \chi_c=\chi_d,\ \chi_a+\chi_b>\chi_d\
\chi_a+\chi_c>\chi_d,\ \chi_b+\chi_c>\chi_d\\
\{\bar{a},\bar{b}\}=\{\bar{b},\bar{c}\}=
\{\bar{c},\bar{a}\}=-\frac{\bar{a}\bar{b}\bar{c}}{\bar{d}},\ \ \\
\{\bar{a},\bar{d}\}=\bar{a}\bar{b}-\bar{a}\bar{c},
\ \ \{\bar{b},\bar{d}\}=\bar{b}\bar{c}-\bar{a}\bar{b},
\ \ \{\bar{c},\bar{d}\}=\bar{a}\bar{c}-\bar{c}\bar{b}
\end{array}
\right.
\ee
\be
7:
\left\{
\begin{array}{l}
\chi_a=\chi_d,\ \chi_b=0,\ \chi_c=\chi_d,\
\{\bar{a},\bar{b}\}=\{\bar{b},\bar{c}\}=
\bar{d}-\frac{\bar{a}\bar{b}\bar{c}}{\bar{d}},\ \ \ \ \ \ \ \ \ \ \ \ \ \
\ \ \ \ \ \ \ \ \ \ \
\\
\{\bar{c},\bar{a}\}=0,\
\{\bar{a},\bar{d}\}=\{\bar{d},\bar{c}\}=-\bar{a}\bar{c},
\ \ \{\bar{b},\bar{d}\}=\bar{b}\bar{c}-\bar{a}\bar{b}
\end{array}
\right.
\ee
\be
8:
\left\{
\begin{array}{l}
0<\chi_a<\chi_d,\ 0<\chi_b<\chi_d,\ \chi_c=\chi_d,\
\chi_a+\chi_b=\chi_d,\
\{\bar{a},\bar{b}\}=
\bar{d}-\frac{\bar{a}\bar{b}\bar{c}}{\bar{d}},\ \ \ \\
\{\bar{b},\bar{c}\}=\{\bar{c},\bar{a}\}=0,\
\{\bar{a},\bar{d}\}=-\bar{a}\bar{c},
\ \ \{\bar{b},\bar{d}\}=\bar{b}\bar{c},
\ \ \{\bar{c},\bar{d}\}=0
\end{array}
\right.
\ee

Let us note that the spectral curve unites $h_{3,2}$
and the cubic Casimir $h_{3,3}$ in the limit $\omega_2\rightarrow\infty$
if $|\frac{x_n-x}{\omega_2}|\neq 1$.

\section{Classification of Limits in $sl(3,\bf C)$ECM}
The classification consists of possible limits of the
hamiltonians $(32-36)$ and the limits of the brackets $(31)$.

A notable fact about classification of the limits of the brackets
takes place if one require non triviality of all three terms in
the quadratic hamiltonians in the limit. In this case $\chi_a$,
$\chi_b$ and $\chi_c$ satisfy some equality coming from
$f_{12}+f_{23}+f_{31}=0$. In the domain $\{I\}$ on fig.$1$ it
looks like $\chi_a+\chi_b+\chi_c=2$ and in $\{III\}$:
$\chi_a=\chi_b+\chi_c$. These equations can be either compatible or not
with the relations determining some given limit of the brackets.
By comparing them with the list of the limits of the brackets given in
Section $5$ we arrive to the fact all types of brackets $(52)-(59)$ are
distributed on fig.$3$ with the midlines drawn on fig.$1$.
 The domains $\{II\}$,
$\{III\}$ and $\{IV\}$ correspond to brackets $8$ - formula $(59)$;
midlines - type $8$, $(60)$ with $\chi_d=1$; the sides of the fundamental
triangle -type $7$, $(59)$; the dual lattice figured on fig.$2$ split
the domain $\{I\}$ on fig.$1$ by three parts: they correspond to the
type $1$ and $4$, $(52),(55)$;  and one dimensional boundaries between them -
type $2$ and $5$, see $(53),(56)$ and the center of the fundamental
triangle - type $3$ and $6$, see $(54),(57)$.
\paragraph{Examples}
Let us write down three examples corresponding to the cases $1,7,8$
($(52),(58),(59)$ correspondingly) for the limits
of the brackets.
\vskip 2mm
Example $1$
\be
\{\bar{a},\bar{b}\}=\bar{d},\ \{\bar{b},\bar{c}\}=0,\ \{\bar{c},\bar{a}\}=0
,\ \{\bar{a},\bar{d}\}=0,\ \ \{\bar{b},\bar{d}\}=0, \ \
\{\bar{c},\bar{d}\}=0\\ f_{12}=\frac{1}{3},\ f_{23}=\frac{1}{4},\
f_{31}=-\frac{7}{12},\ \chi_{a}=\frac{1}{3},\ \chi_{b}=\frac{1}{4},\
\chi_{c}=\frac{1}{2},\ \chi_{d}=\frac{7}{12}\\
\bar{h}_{2,0}=v_1^2+v_2^2+v_3^2-2\bar{a}e^{-U_{12}}-2\bar{b}e^{-U_{23}}
\ \ \ \ \
\bar{h}_{2,2}=2\bar{c}\\
\bar{h}_{3,0}=v_1^3+v_2^3+v_3^3+3v_3\bar{a}e^{-U_{12}}+
3v_1\bar{b}e^{-U_{23}}-3\bar{d}e^{U_{31}}\\
\overline{h_{3,2}+h_{3,3}}=
-3\bar{c}v_2-3\frac{\bar{a}\bar{b}\bar{c}}{\bar{d}},
\ \ \ \ \
\bar{h}_{3,3}=-\frac{3}{2}\bar{d}
\ee
      Example $2$
\be
\{\bar{a},\bar{b}\}=\{\bar{b},\bar{c}\}=
\bar{d}-\frac{\bar{a}\bar{b}\bar{c}}{\bar{d}},\
\{\bar{c},\bar{a}\}=0,\
\{\bar{a},\bar{d}\}=-\bar{a}\bar{c},
\ \ \{\bar{b},\bar{d}\}=\bar{b}\bar{c}-\bar{a}\bar{b},
\ \ \{\bar{c},\bar{d}\}=\bar{a}\bar{c}\\
f_{12}=1,\ f_{23}=0,\ f_{31}=-1,\
\chi_{a}=1,\ \chi_{b}=0,\ \chi_{c}=1,\ \chi_{d}=1\\
\bar{h}_{2,0}=v_1^2+v_2^2+v_3^2-4\bar{a}\cosh(U_{12})-
\frac{1}{2}\bar{b}\frac{1}{\sinh^2(\frac{U_{23}}{2})}-
4\bar{c}\cosh(U_{31}),\ \
\bar{h}_{2,2}=2(\bar{a}+\bar{c})\\
\bar{h}_{3,0}=v_1^3+v_2^3+v_3^3+6v_3\bar{a}\cosh(U_{12})+
\frac{3}{2}v_1\bar{b}\frac{1}{\sinh^2(\frac{U_{23}}{2})}+
6v_2\bar{c}\cosh(U_{31})-\\
-3(\bar{d}-\frac{\bar{a}\bar{b}\bar{c}}{\bar{d}})
\{\sinh(U_{12})+\cosh(U_{12})\coth(\frac{U_{23}}{2})\}\\
\bar{h}_{3,2}=3(v_1+v_2)\bar{a}+3(v_1+v_3)\bar{c}
+\frac{3}{2}(\bar{d}-\frac{\bar{a}\bar{b}\bar{c}}{\bar{d}})
\coth(\frac{U_{23}}{2}),\ \
\bar{h}_{3,3}=-\frac{3}{2}(\bar{d}+\frac{\bar{a}\bar{b}\bar{c}}{\bar{d}})
\ee
      Example $3$
\be
\{\bar{a},\bar{b}\}=
\bar{d}-\frac{\bar{a}\bar{b}\bar{c}}{\bar{d}},\
\{\bar{b},\bar{c}\}=\{\bar{c},\bar{a}\}=0,\
\{\bar{a},\bar{d}\}=-\bar{a}\bar{c},
\ \ \{\bar{b},\bar{d}\}=\bar{b}\bar{c},
\ \ \{\bar{c},\bar{d}\}=0\\
f_{12}=\frac{1}{2},\ f_{23}=\frac{1}{2},\ f_{31}=-1,\
\chi_{a}=\frac{1}{2},\ \chi_{b}=\frac{1}{2},\ \chi_{c}=\chi_{d}=1\\
\bar{h}_{2,0}=v_1^2+v_2^2+v_3^2-2\bar{a}e^{-U_{12}}-2\bar{b}e^{-U_{23}}
-4\bar{c}\cosh(U_{31})\\
\bar{h}_{2,2}=2\bar{c}\\
\bar{h}_{3,0}=v_1^3+v_2^3+v_3^3+3v_3\bar{a}e^{-U_{12}}+
3v_1\bar{b}e^{-U_{23}}+6v_2\bar{c}\cosh(U_{31})
-3(\bar{d}-\frac{\bar{a}\bar{b}\bar{c}}{\bar{d}})e^{U_{31}}\\
\bar{h}_{3,2}=3(v_1+v_3)\bar{c}+\frac{3}{2}
(\bar{d}-\frac{\bar{a}\bar{b}\bar{c}}{\bar{d}}),\ \ \ \
\bar{h}_{3,3}=-\frac{3}{2}(\bar{d}+\frac{\bar{a}\bar{b}\bar{c}}{\bar{d}})\\
\ee

\section{Limits in $sl(2,\bf C)$ EG with two marked points on elliptic
curve.}
Let us consider the following Lax matrix:
\be
L_{kl}=v_k\delta_{kl}+(p_1)_{kk}E_1(z-x_1)+(p_2)_{kk}E_1(z-x_2)+\\
+(p_1)_{kl}e^{2\pi i
\frac{(z-x_1)-(\bar{z}-\bar{x_1 })}{\tau-\bar{\tau}}u_{kl}}
\Phi(u_{kl},z-x_1)+(p_2)_{kl}e^{2\pi i
\frac{(z-x_2)-(\bar{z}-\bar{x_2})}{\tau-\bar{\tau}}u_{kl}}
\Phi(u_{kl},z-x_2)
\ee
\be
\left\{
\begin{array}{c}
v_1=-v_2\\
(p_1)_{11}=-(p_1)_{22}\\
(p_2)_{11}=-(p_2)_{22}
\end{array}
\right.
\ee
Due to $(18)$ we have
\be
\left\{
\begin{array}{c}
(p_1)_{11}=-(p_2)_{11}\\
(p_1)_{22}=-(p_2)_{22}
\end{array}
\right.
\ee
Let us use the following notations
$v=v_1$, $u=u_{12}$ and $p_{11}=(p_1)_{11}$
Instead of the fixing of the action by diagonal $SL(2,\bf C)$ matrix
we again choose invariant coordinates:
\be
(p_1)_{12}(p_1)_{21}=a_1\ \ \
(p_2)_{12}(p_2)_{21}=a_2\ \ \
(p_1)_{12}(p_2)_{21}=a_3\ \ \
(p_1)_{11}=a_4
\ee
The brackets are easily computed:
\be
\{a_1,a_2\}=\{a_1,a_4\}=\{a_2,a_4\}=0\\
\{a_1,a_3\}=\{a_2,a_3\}=2a_3a_4\\
\{a_3,a_4\}=a_3
\ee
To compute the hamiltonians produced by the $tr(L^2)$ we need
the following identity:
\be
\Phi(w,z_1)\Phi(-w,z_2)=-\Phi(w,z_1-z_2)
[E_1(z_1)-E_1(z_2)+E_1(w)-E_1(w+z_1-z_2)]
\ee
The hamiltonians are:
\be
\frac{1}{2}h_{2,0}=v^2-a_1E_2(u)-a_2E_2(u)+\\+a_3e^{2\pi i
\frac{(\bar{x}_1-x_1)-(\bar{x}_2-x_2)}{\tau-\bar{\tau}}u}\Phi(u,x_2-x_1)
[E_1(u+x_2-x_1)-E_1(u)]+\\
+\frac{a_1a_2}{a_3}e^{2\pi i
\frac{(\bar{x}_2-x_2)-(\bar{x}_1-x_1)}{\tau-\bar{\tau}}u}\Phi(u,x_1-x_2)
[E_1(u+x_1-x_2)-E_1(u)],
\ee
\be
\frac{1}{2}h_{2,1,1}=2a_4v-2a_4^2E_1(x_1-x_2)-
a_3e^{2\pi i
\frac{(\bar{x}_1-x_1)-(\bar{x}_2-x_2)}{\tau-\bar{\tau}}u}\Phi(u,x_2-x_1)+\\
+\frac{a_1a_2}{a_3}e^{2\pi i
\frac{(\bar{x}_2-x_2)-(\bar{x}_1-x_1)}{\tau-\bar{\tau}}u}\Phi(u,x_1-x_2),
\\
\frac{1}{2}h_{2,2,1}=a_1+a_4^2,
\ \ \ \
\frac{1}{2}h_{2,2,2}=a_2+a_4^2.
\ee
The notable features of the EG case are:

1.There is non trivial non diagonal part in the matrix $p_{ii}$
that  makes hamiltonians and algebra more complicated.

2.There is a parameter $x_1-x_2$ which can be also shifted.

3.The resultant hamiltonians directly depend on the parameters
of the limit:$f_{ij}$,...

\section{Limits of Lax matrix.}
Using formulas from the Appendix C one can obtain the approximate
expression for the Lax matrix $(6)$, $N_0=1$, $p_{ii}=0$:

\be
L_{kl}=v_k\delta_{kl}-e^{2\pi i
\frac{(z)-\bar{z}}{\tau-\bar{\tau}}u_{kl}}p_{kl}
\Phi(u_{kl},z)
\ee

It degenerates into the following one:

\be
L_{ij}\approx v_{i}+\\+e^{-\frac{U_{ij}}{2}\sigma-f_{ij}
\frac{z_n+\bar{z_n}}{4}-\frac{f_{ij}}{2}\sigma\omega_2}\bar{p}_{ij}
e^{\omega_2\zeta_{ij}}\frac{1}{2}
(\coth\frac{u_{ij}}{2}+\coth\frac{z}{2})(1-e^{u_{ij}+z-2\omega_2})
\approx\\
\approx v_i+\\+
\left\{
\begin{array}{l}
f_{ij}>0:
e^{-\frac{U_{ij}}{2}\sigma-f_{ij}
\frac{z_n+\bar{z_n}}{4}}\bar{p}_{ij}e^{-\frac{f_{ij}}{2}\sigma\omega_2+
\omega_2\zeta_{ij}}
(1-e^{U_{ij}+f_{ij}\omega_2+z_n+\sigma\omega_2-2\omega_2})
\\f_{ij}<0:
e^{-\frac{U_{ij}}{2}\sigma-f_{ij}
\frac{z_n+\bar{z_n}}{4}}\bar{p}_{ij}e^{-\frac{f_{ij}}{2}\sigma\omega_{2}
+\omega_2\zeta_{ij}}(e^{-z_n-\sigma\omega_2}-e^{U_{ij}+f_{ij}\omega_2})
\end{array}
\right.
\ee

The simplest way to obtain the above formula
(besides of the straightforward evaluations given in the Appendix C)
is to use series representation:

\be
\Phi(u,z)=\sum\limits_{n\in \bf Z}\frac
{e^{-nz}}{1-e^{-2n\omega_2-u}}
\ee
Let us consider possible limits.

\paragraph{1. }For $f_{ij}>0$
the requirement of non singularity of $L_{ij}$
when $\omega_2\rightarrow\infty$ is the following:
\be
\left\{
\begin{array}{l}
-\frac{f_{ij}}{2}\sigma+\zeta_{ij}\leq 0\\
-\frac{f_{ij}}{2}\sigma+\zeta_{ij}+f_{ij}+\sigma-2\leq 0
\end{array}
\right.
\ee
Three exists three non trivial limits for the non diagonal part of
$L_{ij}$:
\be
L_{ij}\approx v_i+
\left\{
\begin{array}{l}
1:e^{-\frac{U_{ij}}{2}\sigma-f_{ij}
\frac{z_n+\bar{z_n}}{4}}\bar{p}_{ij}\\
2:e^{-\frac{U_{ij}}{2}\sigma-f_{ij}
\frac{z_n+\bar{z_n}}{4}}\bar{p}_{ij}(-e^{U_{ij}+z_n})\\
3:e^{-\frac{U_{ij}}{2}\sigma-f_{ij}
\frac{z_n+\bar{z_n}}{4}}\bar{p}_{ij}(1-e^{U_{ij}+z_n})
\end{array}
\right.
\ee
The cases $1,2$ and $3$ correspond the following conditions
\be
\left\{
\begin{array}{l}
1:\ f_{ij}=\frac{2}{\sigma}\zeta_{ij},\ \
f_{ij}<2-\sigma,\ \
\zeta_{ij}<\sigma(1-\frac{\sigma}{2}).
\\
2:\ f_{ij}=2-\frac{\zeta_{ij}}{1-\frac{\sigma}{2}},\ \
f_{ij}>2-\sigma,\ \
\zeta_{ij}<\sigma(1-\frac{\sigma}{2}).
\\
3:\ f_{ij}=2-\sigma,\ \
\zeta_{ij}=\sigma(1-\frac{\sigma}{2}).
\end{array}
\right.
\ee
\paragraph{2.}
For $f_{ij}<0$
we have the following non singularity condition for $L_{ij}$:
\be
\left\{
\begin{array}{l}
-\frac{f_{ij}}{2}\sigma+\zeta_{ij}-\sigma\leq 0\\
-\frac{f_{ij}}{2}\sigma+\zeta_{ij}+f_{ij}\leq 0
\end{array}
\right.
\ee
\be
L_{ij}\approx v_i+
\left\{
\begin{array}{l}
4:e^{-\frac{U_{ij}}{2}\sigma-f_{ij}
\frac{z_n+\bar{z_n}}{4}}\bar{p}_{ij}(e^{-z_n})\\
5:e^{-\frac{U_{ij}}{2}\sigma-f_{ij}
\frac{z_n+\bar{z_n}}{4}}\bar{p}_{ij}(-e^{U_{ij}})\\
6:e^{-\frac{U_{ij}}{2}\sigma-f_{ij}
\frac{z_n+\bar{z_n}}{4}}\bar{p}_{ij}(e^{-z_n}-e^{U_{ij}})
\end{array}
\right.
\ee
The cases $4,5$ and $6$ correspond to the following conditions:
\be
\left\{
\begin{array}{l}
4:\
f_{ij}=\frac{2}{\sigma}\zeta_{ij}-2,\ \
f_{ij}<-\sigma,\ \
\zeta_{ij}<\sigma(1-\frac{\sigma}{2}).
\\
5:\
f_{ij}=-\zeta_{ij}\frac{1}{1-\frac{\sigma}{2}},\ \
f_{ij}>-\sigma,\ \
\zeta_{ij}<\sigma(1-\frac{\sigma}{2}).
\\
6:\ f_{ij}=-\sigma,\ \
\zeta_{ij}=\sigma(1-\frac{\sigma}{2}).
\end{array}
\right.
\ee

\paragraph{3. Example}
With the help of the obtained formulas it is possible to obtain the Lax
matrices for the examples given in Section $6$. For the example $3$
(see $(62)$) we have

\be
L(z)=
\left(
\begin{array}{l}
 \ \ \ \ \ \ \ \ \ v_1 \ \ \ \ \ \ \ \ \ \ \ \ \ \
\ \ -\bar{p}_{12}e^{-\frac{1}{2}U_{12}}
\ \ \ \ \ \bar{p}_{13}e^{-\frac{1}{2}U_{31}}(e^{U_{31}}-e^{-z})\\
\ \ \ \ \ \ \ \ \bar{p}_{21}e^{-\frac{1}{2}U_{12}}
\ \ \ \ \ \ \ \ \ \ \ \ \ \ \ \ \ v_2
\ \ \ \ \ \ \ \ \ \ \ \ \ \ \ \ \ -\bar{p}_{23}e^{-\frac{1}{2}U_{23}}\\
-\bar{p}_{31}e^{\frac{1}{2}U_{31}}
(1-e^{-U_{31}}e^{z})\ \ \ \ \ \ \ \bar{p}_{32}e^{-\frac{1}{2}U_{23}}\ \ \ \
\ \ \ \ \ \ \ \ \ \ \ \ \ \ v_{3}
\end{array}
\right)
\ee

\section{Conclusion.}

In the recent paper it is shown that the IL applied to
the $sl(N,\bf C)$ elliptic Euler-Calogero Model
and the elliptic Gaudin Model
produces new Toda-like systems of $N$ interacting particles
endowed with additional degrees of freedom corresponding to a
coadjoint orbit in $sl(n,\bf C)$. The limits corresponding
to the complete degeneration of the orbital degrees provide only
ordinary periodic and non periodic Toda systems.
We introduce a classification of the systems appearing from
the $sl(3,\bf C)$ case.
The classification is represented on two-dimensional
space of parameters describing the infinite shifts of the coordinates.
This space is subdivided into symmetric domains.
The mixture of the Toda and the Sutherland-Calogero potentials
emerges on the low dimensional domain walls of this picture.
Due to obvious symmetries this classification can be generalized to
the arbitrary number of particles.
We also apply the Inozemtsev limit to
$sl(2,\bf C)$ elliptic Gaudin Model with two marked points on the elliptic
curve and discuss main features of its possible limits.
The limits of Lax matrices are also considered.

We are grateful to A.Chervov, A.Marshakov, A.Gorsky, A.Mironov, A.Morozov
and A.Zabrodin for useful remarks and to A.Levin for many discussions.
We especially grateful to M.A.Olshanetsky for initiating the
work and useful discussions.
The work of both authors was partially supported by RFBR grant
N00-02-16530 and the program for support of the scientific
schools 00-15-96557.

\section{Appendix A: Basic Formulas and Definitions}
The rest of the formulas given in appendices are borrowed
from [5],[8] and [10].

The basic element is the theta function:

\be
\vartheta(z,\tau)=q^{\frac{1}{8}}e^{-\frac{\pi}{4}}(e^{i\pi z}-e^{-i\pi z})
\prod\limits_{n=1}^{\infty}(1-q^{n})(1-q^{n}e^{2\pi iz})
(1-q^n e^{-2\pi i z}),\ \ q=e^{2\pi i \tau}
\ee
The series representation for the Eisenstein functions is convenient
to make the limit $\omega_2\rightarrow\infty$:
\be
E_1(z,\tau)=\partial_{z}log\vartheta(z,\tau),\ \
E_1(z,\tau)\approx \frac{1}{z}+...
\ee

\be
E_2(z,\tau)=-\partial_z E_1(z,\tau),\ \
E_2(z,\tau)\approx \frac{1}{z^2}+...
\ee
Relations to the Weierstrass functions:
\be
\zeta(z,\tau)=E_1(z,\tau)+2\eta_1(\tau)z
\ee

\be
\wp(z,\tau)=E_2(z,\tau)-2\eta_1(\tau)
\ee
where
$$
\eta_1(\tau)=\zeta(\frac{1}{2})
$$
In the Appendices B and C we use notations which are
convenient for making the limits (see [8]):
\be
\left\{
\begin{array}{c}
\tau=\frac{\omega_2}{\omega_1},\\
\omega_1=-i\pi\ \ \ Im(\omega_2)=0
\end{array}
\right.
\ee
\section{Appendix B: formulas which are used for computing
the hamiltonians and their limits}
Let us represent the Eisenstein functions in the form convenient
to make the limit $\omega_2\rightarrow\infty$:
\be
E_1(z)=\frac{1}{2}\sum\limits_{k=-\infty}^
{\infty}\coth(\frac{z}{2}-k\omega_2)
\ee

\be
E_2(z)=\frac{1}{4}\sum\limits_{k=-\infty}^{\infty}
\frac{1}{\sinh^2(\frac{z}{2}-k\omega_2)}
\ee

\be
E_2'(z)=-\frac{1}{4}\sum\limits_{k=-\infty}^{\infty}
\frac{\cosh(\frac{z}{2}-k\omega_2)}{\sinh^3(\frac{z}{2}-k\omega_2)}
\ee

The following important expression appears in the Lax matrix:
\be
\Phi(u,z)=\frac{\vartheta(u+z)\vartheta'(0)}{\vartheta(u)\vartheta(z)}
\ee
It has a pole at $z=0$ and
$$
res|_{z=0}\Phi(u,z)=1
$$
The well known identities are:
\be
\Phi(u,v)\Phi(-u,v)=E_2(v)-E_2(u)
\ee

\be
\Phi'_u(u,v)=\Phi(u,v)(E_1(u+v)-E_1(u))
\ee

And consequently
\be
\frac{E_2'(u)}{E_2(u)-E_2(v)}=E_1(u+v)+E_1(u-v)-2E_1(u)
\ee
The following identity is necessary to find the cubic hamiltonians:
\be
det
\left(
\begin{array}{c}
1\ E_2(u)\ E_2'(u)\\
1\ E_2(v)\ E_2'(v)\\
1\ E_2(w)\ E_2'(w)
\end{array}
\right)=\frac{2\vartheta(v-w)\vartheta(w-u)\vartheta(u-v)\vartheta(u+v+w)}
{[\vartheta(u)\vartheta(v)\vartheta(w)]^3}
\ee
It can be rewritten in the following way:
\be
\Phi(u,z)\Phi(v,z)\Phi(-u-v,z)=-\frac{1}{2}E_2'(z)-\frac{1}{2}E_2(z)
\frac{E_2'(u)-E_2'(v)}{E_2(u)-E_2(v)}-\\
-\frac{1}{2}\frac{E_2'(v)E_2(u)-E_2'(u)E_2(v)}{E_2(u)-E_2(v)}
\ee
Using formula $(93)$ we have
\be
\frac{E_2'(u)-E_2'(v)}{E_2(u)-E_2(v)}=2E_1(u+v)-2E_1(u)-2E_1(v)
\ee
\be
\frac{E_2'(v)E_2(u)-E_2'(u)E_2(v)}{E_2(u)-E_2(v)}
=-E_1(u+v)(E_2(u)+E_2(v))+\\+E_1(u-v)(E_2(u)-E_2(v))+2(E_2(u)E_1(v)
+E_2(v)E_1(u))
\ee
Thus formula $(95)$ can be written in the following form:
\be
\Phi(u,z)\Phi(v,z)\Phi(-u-v,z)=-\frac{1}{2}E_2'(z)+E_2(z)
[E_1(u)+E_1(v)+E_1(-u-v)]-\\-E_2(u)E_1(v)-E_2(v)E_1(u)+\\+
\frac{1}{2}E_1(u+v)(E_2(u)+E_2(v))-\frac{1}{2}E_1(u-v)(E_2(u)-E_2(v))
\ee
\section{Appendix C: formulas which are used for computing the limits
in the Lax matrix}
\be
\vartheta(z,\tau)=q^{\frac{1}{8}}e^{-\frac{\pi}{4}}(e^{i\pi z}-e^{-i\pi z})
\prod\limits_{n=1}^{\infty}(1-q^{n})(1-q^{n}e^{2\pi iz})
(1-q^n e^{-2\pi i z}),\ \ q=e^{2\pi i \tau}
\ee
\be
\sum\limits_{n=1}^{\infty}1=\zeta(0)=-\frac{1}{2}\ \ \ \ \ \ \
\sum\limits_{n=1}^{\infty}n=\zeta(-1)=-\frac{1}{12}
\ee
where $\zeta(z)$ is the Riemann zeta function.
\be
\frac{\vartheta(u,\tau)}{\vartheta'(0,\tau)}=\frac{\omega_1}{\pi}
q^{-\frac{1}{12}}\prod\limits_{n=-\infty}^{\infty}\sin(\pi(u-n\tau))
\ee
\be
\frac{\vartheta(\frac{z}{2\pi i},\tau)}{\vartheta'(0,\tau)}=
-ie^{\frac{\omega_2}{6}}\prod\limits_{n=-\infty}^{\infty}
\sinh(\frac{z}{2}+n\omega_2)=
\\
=-2ie^{\frac{\omega_2}{6}}
\sinh\frac{z}{2}
\prod\limits_{n=1}^{\infty}(e^z+e^{-z}-e^{2n\omega_2}-e^{-2n\omega_2})=
\\
=-2i\cdot \sinh{\frac{z}{2}}\prod\limits_{n=1}^{\infty}(-1+e^{z-2n\omega_2}
+e^{-z-2n\omega_2}-e^{-4n\omega_2})\approx
\\
\approx 2i\cdot
\sinh(\frac{z}{2})(1-e^{z-2\omega_2}-e^{-z-2\omega_2})
\ee
If $\frac{\partial z}{\partial \omega_2}>0$ and
$\frac{\partial u}{\partial \omega_2}>0$ than
\be
\frac{(1-e^{u+z-2\omega_2}-e^{-u-z-2\omega_2})}
{(1-e^{u-2\omega_2}-e^{-u-2\omega_2})(1-e^{z-2\omega_2}-e^{-z-2\omega_2})}
\approx 1-e^{u+z-2\omega_2}
\ee

\end{document}